\documentclass[aps,preprint]{revtex4}
\usepackage{times}

\usepackage{graphicx}

\newcommand{\be}{\begin{equation}}
\newcommand{\ee}{\end{equation}}
\newcommand{\bea}{ \begin{eqnarray}}
\newcommand{\eea}{ \end{eqnarray}}

\newcommand{\Eq}[1]{Eq.~(\ref{#1})}
\renewcommand{\dag}{^{\dagger}}
\newcommand{\la}{\left<}
\newcommand{\ra}{\right>}

\renewcommand{\r}{\mathbf{r}}
\newcommand{\R}{\mathbf{R}}
\renewcommand{\k}{\mathbf{k}}
\newcommand{\q}{\mathbf{q}}
\newcommand{\re}{\mathbf{r}_e}
\newcommand{\rh}{\mathbf{r}_h}

\newcommand{\ek}{\epsilon_\k}

\begin{document}

%\date{today}
%\keywords{Exciton, Bose-Einstein condensation, T-matrix, coupled
%quantum wells.}
%\subjclass[pacs]{71.35.Lk, 03.75.Hh, 78.67.De}

\title{Dynamical T-matrix theory
for high-density excitons\\ in coupled quantum wells}

\author{R. Zimmermann}
\affiliation{Institut f\"ur Physik der Humboldt-Universit\"at zu
Berlin, Newtonstr. 15, D-12489 Berlin, Germany}

\begin{abstract}
Excitons in coupled quantum wells open the possibility to reach
high densities close to equilibrium. In a recent experiment
employing a lateral trap potential, a blue shift and a broadening
of the exciton emission line has been seen \cite{SnokeEx}. The
standard Hartree-Fock treatment can explain the blue shift but
fails to give a finite broadening. Starting from the
(spin-dependent) many-exciton Hamiltonian with direct and
exchange potential, we present a dynamical T-matrix calculation
for the single-exciton Green's function which is directly related
to the frequency- and angle-resolved photoluminescence. The
calculated spectrum is blue shifted and broadened due to
exciton-exciton scattering. At high excitation, both the spectrum
and the angular emission are getting narrow. This is a direct
manifestation for off-diagonal long range order and a precursor
of condensation.
\end{abstract}

\maketitle

\section{Introduction}

On the search for exciton systems where high densities at low
temperatures can be reached, coupled quantum wells (CQW) came
into the focus recently \cite{Butov,Snoke}. Being spatially
indirect, the excitons have microsecond lifetimes, and
equilibration at Helium temperatures can be expected. In
addition, lateral confinement into a stress-induced trap will
enforce high local densities. Standard estimates of critical
temperatures/critical densities for Bose-Einstein condensation
(BEC) gave promising values, thanks to the small exciton mass.
However, excitons in CQW feel a strong dipole-dipole repulsion,
and expressions valid for a nearly ideal Bose gas cannot be
applied. Recent experiments by Snoke and coworkers \cite{SnokeEx}
have shown a substantial blue shift (5\,meV) of the exciton
emission line under high excitation in a trap, but accompanied by
a sizable broadening of the exciton line. For an overview on
attempts to find exciton BEC in semiconductors, see the special
issue on ''Spontaneous Coherence in Excitonic Systems''
\cite{SSC}.

A Hartree-Fock treatment of the many-exciton problem can easily
explain the blue shift. However, while the Hartree-Fock approach
(including a c-number term) is at the heart of the Bogolubov
theory of BEC \cite{Griffin}, it is not able in principle to
describe line broadenings. In order to go one step further, we
present here a dynamical T-matrix theory where multiple
exciton-exciton (XX) scattering is included, and which gives
realistic line shapes. The theory presented is not applicable to
the BEC state itself. However, the approach to condensation is
investigated in more detail than before. Finding characteristic
features in the emission spectrum and its angular dependence
\cite{Keeling} may help to specify conditions in favor of BEC.

\section{Excitons in coupled quantum wells}

\begin{figure}[t]
\includegraphics[angle=-90,width=.4\textwidth]{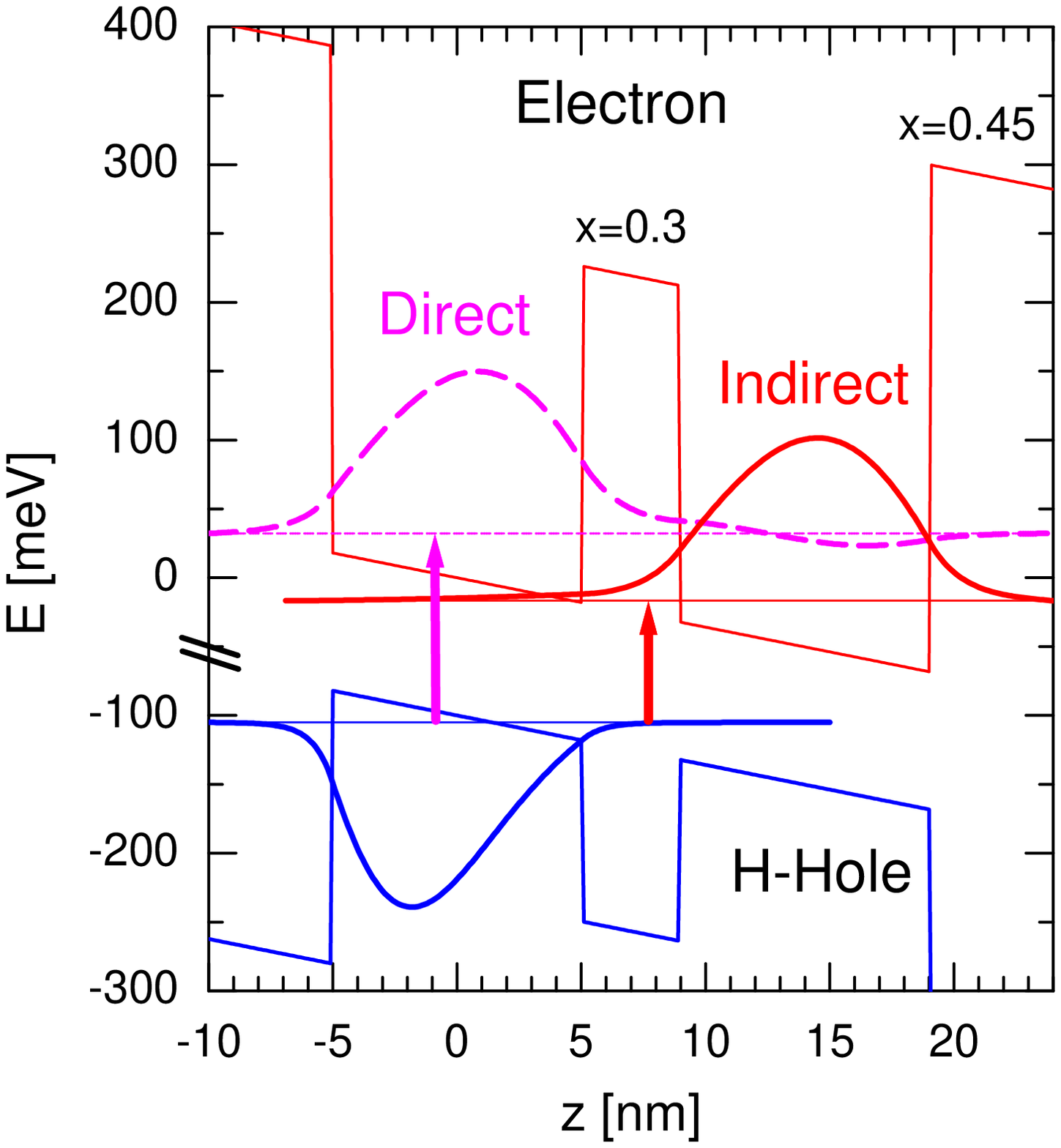}
\caption{Band edge diagram for the coupled quantum well system
used in Ref.~\cite{SnokeEx}: Two GaAs quantum wells of 10\,nm
width are separated by an Al$_x$Ga$_{1-x}$As barrier of 4\,nm. A
static electric field of $F = 36\,$kV/cm is tilting the band
edges. Each confinement wave function is plotted with a vertical
offset being  its energy.} \label{Fig:1}
\end{figure}
Applying a static electric field in the growth (z-) direction
allows to tune the coupled quantum well such that the indirect
exciton state becomes the lowest one (see Fig.~1). The
single-exciton wave function can be factorized, denoting the
in-plane relative coordinate with $\r$ and the center-of-mass one
with $\R$, as
\be \Psi(\re,\rh) = u_e(z_e)\,u_h(z_h)\,\phi(\r)\,\psi(\R) \, .
\ee
The confinement functions $u_a(z_a)$ for the CQW system under
study are shown in Fig.~1, too. The Schr\"odinger equation for
the 1s exciton wave function $\phi(\r)$ is solved numerically
\cite{ZimSSC} with the potential
\be v_{eh}(\r) = \int dz \, dz' \frac{e^2}{4\pi\varepsilon_0
\varepsilon_s \sqrt{r^2+(z-z')^2}} \, u_e^2(z) \, u_h^2(z') \ee
and gives an indirect (direct) exciton binding energy of
3.5\,(19.1)\,meV. The calculated radiative lifetimes are
$\tau_{dir} = 31\,$ps and $\tau_{ind} = 0.45\,\mu$s. Therefore,
equilibration of the indirect excitons at low bath temperatures
may be achieved.

However, at the same time, indirect excitons feel a strong
Coulomb repulsion of dipole-dipole character. Reducing this to a
contact potential gives the strength
\be U_d = \int \! d\r \left[v_{ee}(\r) + v_{hh}(\r) - 2v_{eh}(\r)
\right] \, . \ee
In the strict 2D limit, this would result in the dipole-dipole
repulsion of $U_d = d e^2/\varepsilon_0 \varepsilon_s$, where $d$
is the effective CQW separation. Additionally, the internal
fermionic structure of excitons leads to a (non-local) exchange
potential. Bringing this into contact form as well, we have
\be U_x = \sum_{\k,\k'} \left[2 v_{eh}(\k-\k') \,\phi^3_\k
\,\phi_{\k'} \, -\,(v_{ee}(\k-\k') + v_{hh}(\k-\k'))\, \phi^2_\k
\,\phi^2_{\k'}\right] \, . \ee
For the CQW of Fig.\,1 we have calculated $U_\mathrm{d} =
18.8\,$eV nm$^2$ and $U_\mathrm{x} = -8.9\,$eV nm$^2$.

With these ingredients, and taking into account the spin
structure of excitons composed from spin 1/2 conduction electrons
and spin 3/2 heavy hole states, the following many-exciton boson
Hamiltonian can be constructed \cite{Laikht}
\bea \label{Ham} H & = & \int \! d\R \sum_s \Psi\dag_s(\R) \left[
-\frac{\hbar^2 \nabla^2}{2M} + V(\R)\right] \Psi_s(\R) \,
+ \nonumber\\
& + & \frac{1}{2}(U_d+U_x) \int \! d\R \sum_{s s'}
   \Psi_s\dag(\R) \Psi_{s'}\dag(\R) \Psi_{s'}(\R) \Psi_s(\R)\\
& + & U_x \int \! d\R \sum_{s s'} \left(\frac{1}{4} -
\delta_{ss'}\right)
   \Psi_s\dag(\R) \Psi_{-s}\dag(\R) \Psi_{-s'}(\R) \Psi_{s'}(\R) \, . \nonumber
\eea
The spin label $s$ denotes the four exciton states, which are
$s=\pm 1$ (bright) and $s= \pm 2$ (dark). The one-exciton
potential $V(\R)$ can model a lateral trap confinement, but is
not considered here.

\section{Dynamical T-matrix theory}

In diagram language, we sum up multiple XX scattering events to
form the dynamical T-matrix $T_\q(z)$. Plugging this into the
one-exciton self energy $\Sigma_\k(z)$ leads to an improved
one-exciton propagator $G_\k(z)$, as depicted in Fig.~2. This
selfconsistency cycle has to be repeated up to convergency.
\begin{figure}[t]
\includegraphics[angle=-90,width=.4\textwidth]{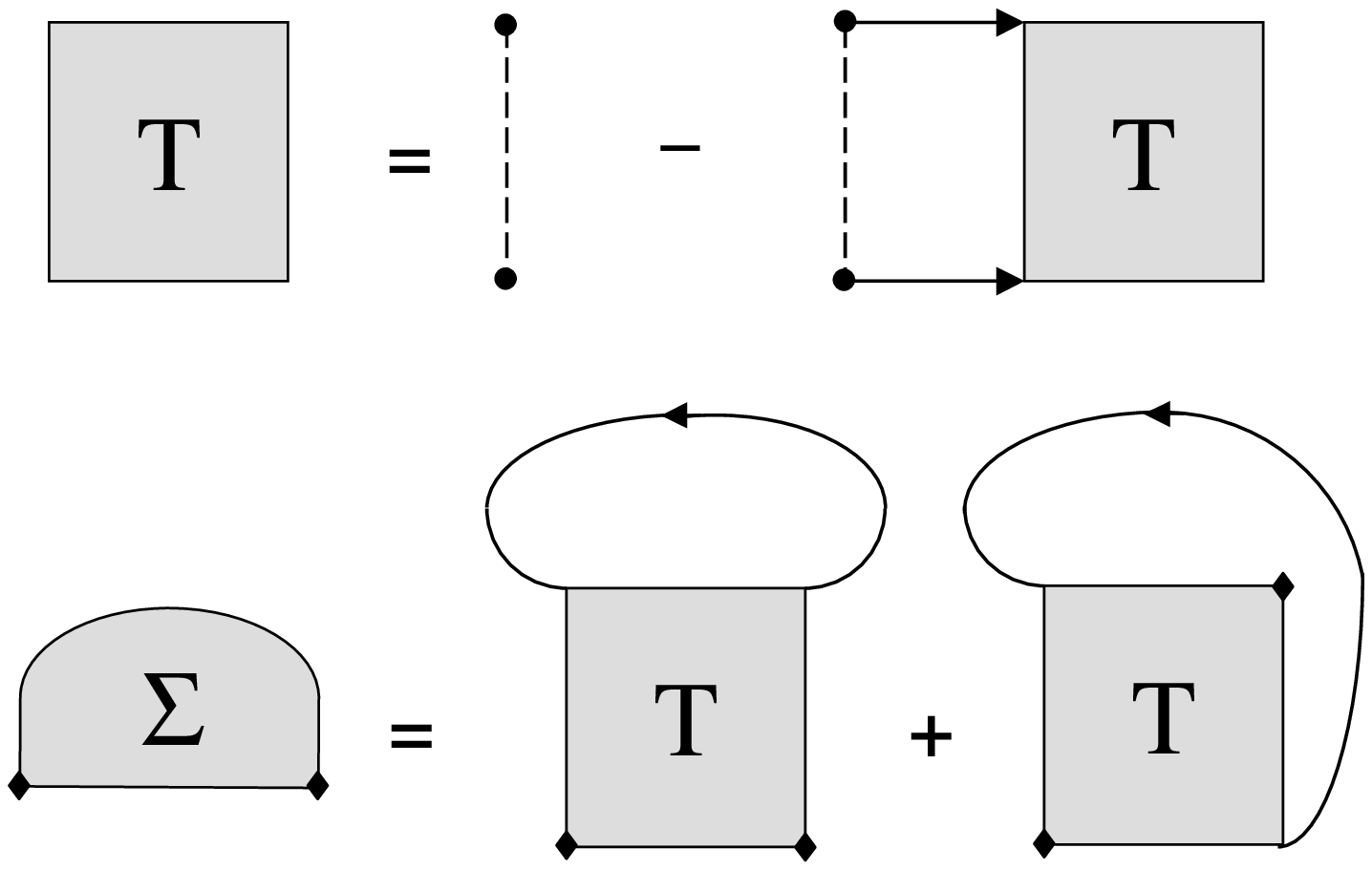}
\caption{Diagrammatic representation of the dynamical T-matrix
scheme. The full line with arrow denotes the single exciton
Green's function $G_\k(z)$, and the dashed line is the XX
interaction $U^\pm$ (for simplicity, we have not resolved the
spin degree of freedom here). The two terms in the self energy
are the direct and (boson) exchange contributions. For the
contact interaction used, they can be combined into one term.}
\label{Fig:2}
\end{figure}
The specific spin structure of \Eq{Ham} allows to split the
T-matrix into a bonding/antibonding part $T^\pm$, which obey
\be T_\q^\pm(\Omega) = \frac{U^\pm}{1\,+\, U^\pm \,{\cal
G}_\q(\Omega)} \, , \quad \quad U^\pm = U_d \, \pm \, U_x \, .\ee
The exciton pair propagator (two parallel arrows in Fig.~2) is
given by
\be \label{PairProp} {\cal G}_\q(\Omega) = \sum_\k \int
\frac{d\omega}{\pi} \frac{d\omega'}{\pi} \,A_\k(\omega)
\,A_{\k-\q}(\omega') \,\frac{1 + 2 g_B(\hbar\omega-\mu)}{\omega
+\omega' - \Omega} \, , \quad  g_B(E) = \frac{1}{\exp(E/k_B T) -
1} \, . \ee
In writing \Eq{PairProp} we have used a fixed chemical potential
$\mu$ in the Bose distribution function $g_B(\hbar\omega - \mu)$,
independent on spin label. Thus, complete spin relaxation of
excitons into equilibrium has been assumed. Taken as it stands,
the real part of \Eq{PairProp} has a logarithmic divergency
coming from the integration over $\k$. This is a well-known
shortcoming of using in two dimensions a contact potential. As a
remedy, we have chosen to cut off the integration at $\ek =
100\,$meV, well above any relevant energy scale.

The self energy follows with $T \equiv (9/2)T^+ \, + \, (1/2)
T^-$ as
\bea \label{SigT} \Sigma_\k(z) &=& \sum_{\q} \int
\frac{d\omega}{\pi}
A_{\k-\q}(\omega) \times \\
&&\times \left[T_\q(z + \omega) \,g_B(\hbar\omega-\mu) \, + \,
\int \frac{d\omega'}{\pi} \,\frac {\mathrm{Im}T_\q(\omega' -
i0)\, g_B(\hbar\omega'-2\mu)}{\omega' - z - \omega} \right]
\nonumber \eea
and enters the exciton Green's function resp. its spectral
function
\be \label{Spec} A_\k(\omega) = \textrm{Im}G_\k(\omega-i0) =
\frac{\mathrm{Im}\Sigma_\k(\omega-i0)} {(\omega-\epsilon_\k -
\mathrm{Re}\Sigma_\k(\omega-i0))^2 \, +\,
(\mathrm{Im}\Sigma_\k(\omega-i0))^2} \, . \ee
For the exciton system, the spectrally and directionally resolved
spontaneous optical emission is given by
\be \label{PL1} I(\k,\omega) = \mu_{cv} \int \! d\R \, d\R' \,
e^{i\k(\R-\R')} \int\! dt \, e^{i\omega t} \, \la
\Psi\dag_s(\R,t) \Psi_s(\R',0)\ra \, , \quad (s = \pm 1) \ee
which shows clearly the importance of contributions $\R \neq \R'$
(off-diagonal long range order)  for the directional characteristic
(dependence on $\k$). Further, \Eq{PL1} can be simply expressed
via the single-exciton propagator resp. the spectral function,
\be I(\k,\omega) \propto i\, G^{<}_\k(\omega) \equiv
 \,g_B(\hbar\omega-\mu)\, A_\k(\omega)\, .\ee
For understanding the spectral shape, it is important to note
that the spectral function changes sign at the chemical potential
$\mu$ where the Bose distribution function $g_B(\hbar\omega -
\mu)$ has a pole, resulting in a strictly positive emission.
Portions below $\mu$ are due to photon emission accompanied by XX
scattering. The classical argument for BEC onset ($\mu$ is
touching the energy of the lowest state) has to be refined here:
A phase transition happens if the chemical potential hits the
quasiparticle dispersion, $\mathrm{Re}\Sigma_0(\hbar\omega=\mu) =
\mu$. Then, both the spectrum and the directional characteristic
evolve into sharp delta peaks. In Fig.~3, results of the full
dynamical T-matrix calculation (at zero momentum) are shown. The
exciton emission increases with rising density. The initial
increase in line width due to XX scattering turns into sharpening
of the spectral line shape (left), while a peak in the angle
resolved emission (right) evolves. However, condensation is not
achieved at $T = 5\,$K for the density range considered.

The exciton densities given in Fig.~3 are calculated with the
standard expression using the momentum- and frequency-dependent
spectral function \Eq{Spec} and the Bose distribution function.
It is worth noting that the T-matrix comes out appreciably
smaller than the bare interaction ($U^\pm$). Consequently, at a
given density, the blue shift of the emission is much less than a
simple Hartree-Fock argument would predict. Obviously, the strong
dipole-dipole repulsion and its dynamical character hinder an
easy build-up of coherence. Further calculations including the
confining action of a lateral trap shall specify more precisely
conditions for condensation in coupled quantum wells.

\begin{figure}[t]
\begin{center}
\includegraphics[angle=-90,width=.38\textwidth]{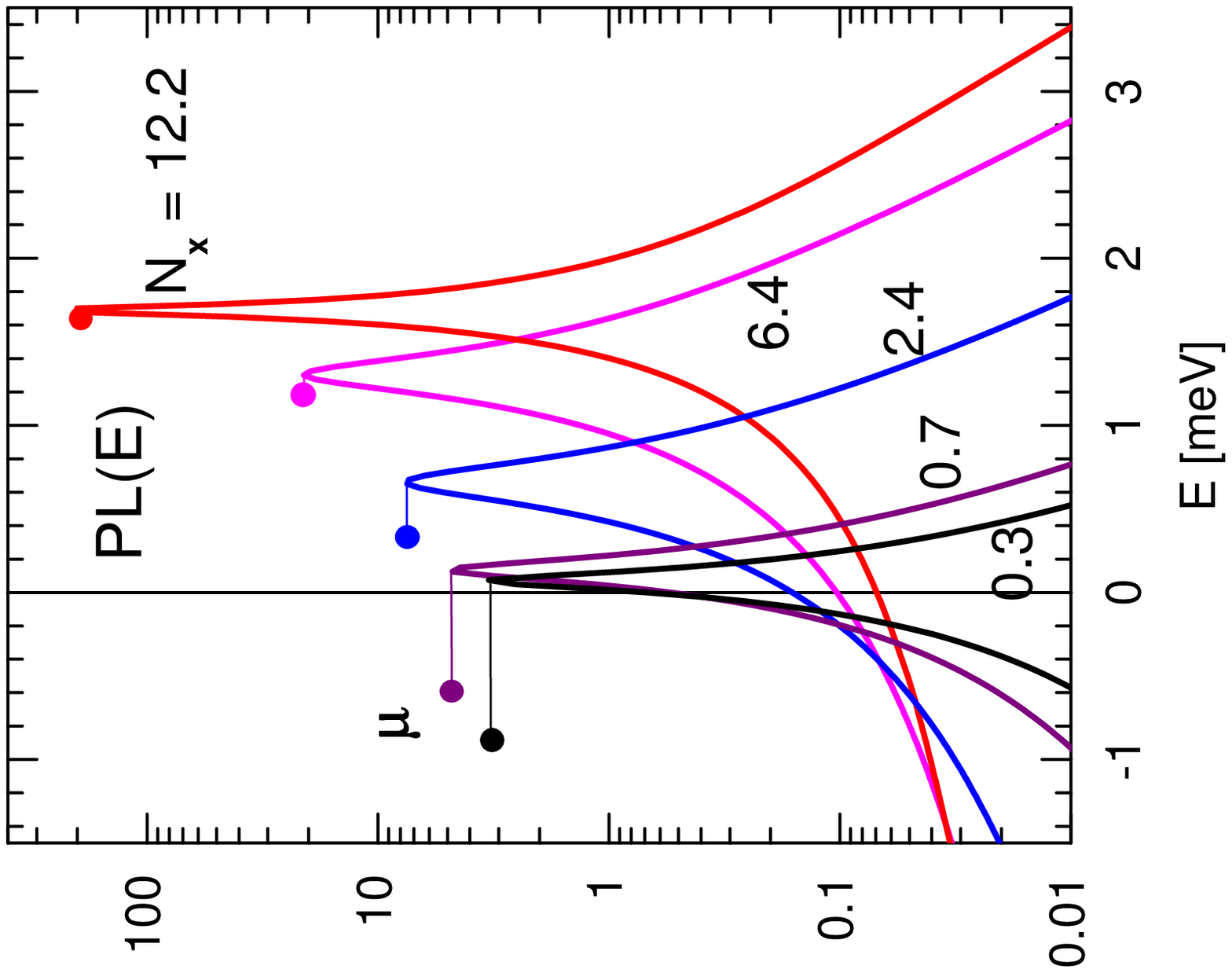}~~~
\includegraphics*[angle=-90,width=.38\textwidth]{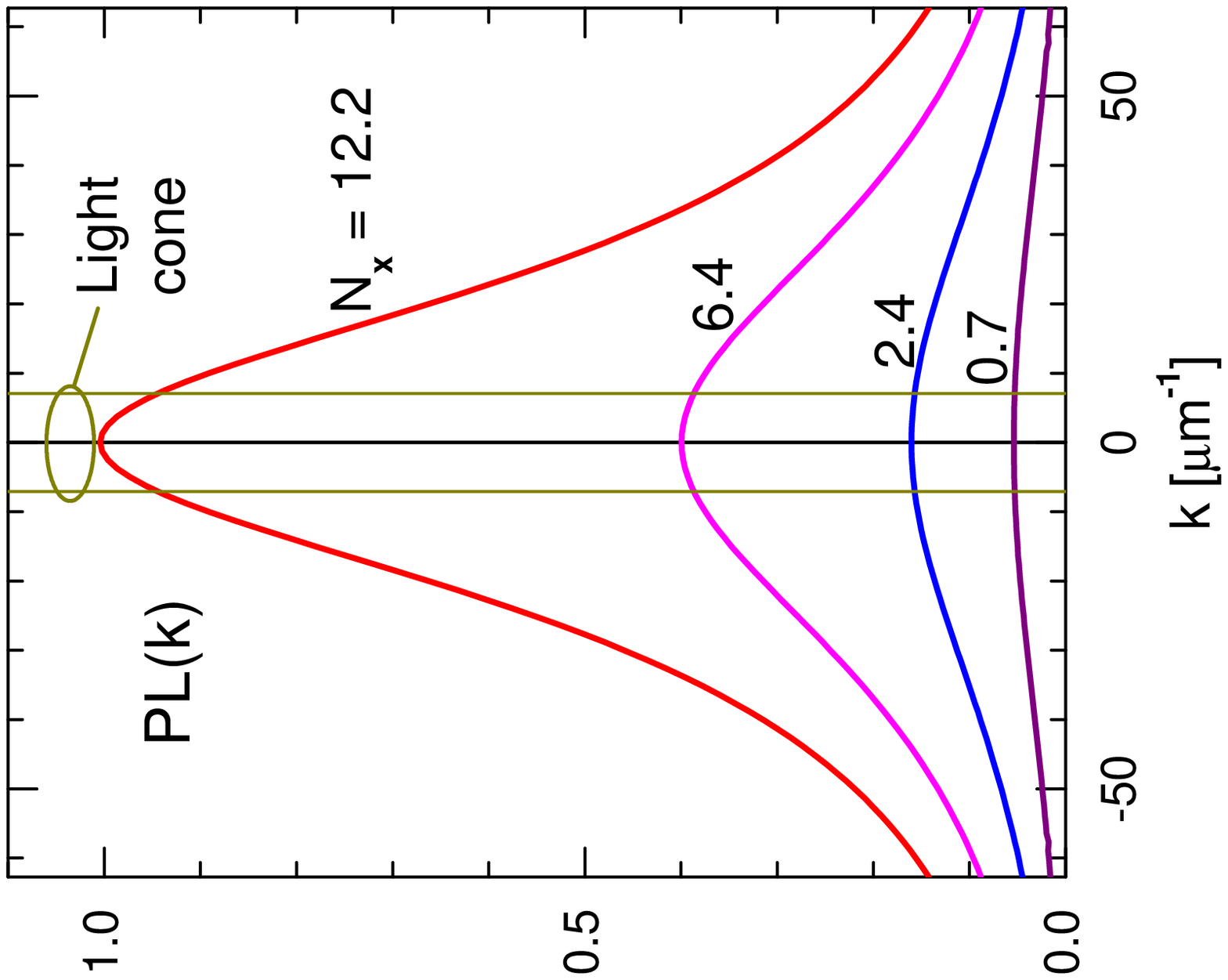}
\end{center}
\caption{Emission line shapes in dependence on frequency (left)
and on direction (right). The lines are getting sharper as the
chemical potential $\mu$ (dots) approaches the emission maximum
(quasiparticle position).  The temperature is held fixed at
$T=5\,$K, and exciton densities $N_X$ are given in units of
$10^{10}\,$cm$^{-2}$. The light cone (marked by vertical lines on
the right) sets a limit for the observability of the angular
emission.} \label{Fig:3}
\end{figure}

\end{document}